# A wind turbine efficiency limit higher than the Lanchester (Betz) limit

Thad S. Morton

A generalized actuator-disk analysis is developed that retains both the axial and tangential velocity components of the flow through the blade passage. Unlike the classical one-dimensional formulation, the present theory distinguishes between the flow immediately upstream and downstream of the rotor, accounts explicitly for the rotation imparted to the flow, and derives turbine performance from the coupled conservation of mass, linear momentum, angular momentum, and energy. Optimization of the resulting power coefficient reveals two distinct solution branches. At lower blade-speed ratios, the optimum lies on the constraint curve defined by recovery of the velocity magnitude at the rotor-exit plane to the freestream value. Above a critical blade-speed ratio, however, the optimum leaves this constraint boundary and follows a second branch characterized by an exit speed below the freestream value. The transition between these branches is obtained analytically as the intersection of the unconstrained optimum and the boundary-constrained optimum, yielding a quartic equation whose physically admissible root determines the transition point. The optimized solutions predict power coefficients that exceed the classical Betz limit (reaching 78% at its peak) while remaining fully consistent with the governing conservation laws. The analysis shows that the classical efficiency limit is a consequence of the one-dimensional flow assumption rather than a fundamental upper limit imposed by the governing conservation laws.

---

## 1. Background

The efficiency of a wind turbine system normally includes the aerodynamic efficiency of the blades, the efficiency of the gearbox, if any, and that of the generator. The aerodynamic efficiency referred to in this study is the fraction of power reaching the rotating shaft divided by the total rate at which energy approaches the turbine in the form of wind. This is sometimes called the power coefficient.

As van Kuik (2007) pointed out, the well-known wind turbine aerodynamic efficiency limit of 59% was first published by Lanchester (1915) and then by Joukowsky (1920) and Betz (1920, 1927), so it is called the Lanchester-Betz-Joukowsky limit (see also Bergey 1979). However, the work of Betz was the most available to readers, so the name “Betz limit” was used for some years. For more on this aspect of the history, see Okulov and van Kuik (2012).

Before the development of modern blade-element momentum methods, swirl was recognized as an important feature of rotating machinery, but it was not generally incorporated as an essential variable in the aerodynamic analysis of ideal rotor performance. Glauert (1926, 1935) combined blade-element concepts with momentum methods and incorporated rotational effects into propeller and rotor analysis. However, the optimal finite-blade problem was addressed separately by Goldstein (1929), who assumed that the optimal wake is a helical free vortex and determined the blade circulation distribution required to produce it. Goldstein derived and tabulated the optimal circulation distribution (termed his circulation factor), providing the optimal loading distribution for finite-bladed rotors. Larrabee (1979) provided an accessible modern treatment of ideal propeller theory based on the work of Goldstein and Glauert and discussed the relationship between induced velocity, swirl, and efficiency.

Goldstein's and Glauert's analyses addressed the question of optimal rotor loading within an assumed rotational flow. The question they examined was: Given the desired wake structure, what blade loading produces the most efficient rotor? The present analysis addresses a different question: Without prescribing the wake structure, what flow field is required by the conservation laws to achieve maximum energy extraction? The resulting limit is therefore an actuator-disk bound rather than an optimal blade-loading solution.

After Hütter (1977) suggested that the limit of 59% might not represent a definitive upper bound, and Inglis (1979) pointed out that at least one wind turbine may have achieved an efficiency higher than this limit, Greet (1980) observed that no such limit can be found in a one-dimensional analysis. Rauh and Seelert (1984) agreed that a theoretically grounded optimum for windmill aerodynamic efficiency had not yet been established.

The Lanchester–Betz limit was developed solely from an axial momentum tradition, but rotor aerodynamic theory subsequently developed into a vector-based approach—one that recognized that torque extraction and swirl are inseparable. The 1-D formulation used to derive the 59% upper limit, often referred to as General Momentum Theory (GMT) or actuator-disk theory, does not explicitly account for swirl. Van Kuick (2020) revisited the actuator-disk efficiency limit and examined the role of velocity conditions at the rotor plane, but retained the classical framework in which the inlet and exit velocities are not treated as independent physical states. The present work contends that the Lanchester–Betz theoretical efficiency limit should have been updated after rotor theory moved from scalar momentum theory to vector rotor aerodynamics.

Blade-element momentum (BEM) methods extend rotor analysis by dividing the blade into elements, each treated as an aerodynamic section that generates lift and drag. These local forces are combined with momentum relations to predict rotor performance. Previous actuator-disk formulations, by contrast, generally treated the rotor as a single flow-control surface. Sharpe (2004) and Wood (2007) extended actuator-disk theory by introducing radial variation in the flow variables, allowing the rotor to be analyzed as a collection of annular stream regions. These extensions provided a more detailed description of radial flow structure but did not seek to determine an optimized induction factor or an upper efficiency limit. The axial induction factor $a$, defined as

$$a = \frac{(v_\infty - v_1)}{v_\infty}. \tag{1}$$

is the amount by which the freestream wind at velocity $v_\infty$ is slowed to $v_1$ at the inlet plane of the wind turbine The Lanchester-Betz derivation yields $a = 1/3$ (see (18) below).

The present formulation builds on the annular actuator-disk viewpoint of Sharpe and Wood but differs fundamentally from BEM. The stream annuli are not treated as independent blade elements requiring prescribed airfoil geometry, lift, drag, or aerodynamic coefficients. Instead, each annulus is governed by the coupled conservation of mass, linear momentum, angular momentum, and energy while explicitly retaining both axial and tangential velocity components. BEM starts with the blade and predicts the flow; the present method starts with the optimal flow and determines the aerodynamic requirements the blade must satisfy.

Goldstein assumed the optimal wake and inferred the blade loading required to produce it. BEM begins with the blade and predicts the flow. The present theory begins with the governing conservation laws and determines the optimal wake, from which the aerodynamic requirements of the blade are inferred. This can be summarized as follows:

| | |
|---|---|
| Goldstein: | wake → blade. |
| BEM: | blade → wake. |
| Present paper: | conservation laws → wake → blade. |

To the author's knowledge, no previous study has proposed an alternative to the classical 59% efficiency limit for a conventional wind turbine. One reason is the common expectation that incorporating swirl would only further reduce efficiency because part of the available energy will be diverted into rotational motion. But this reasoning misses the broader questions: What happens to the axial induction factor when swirl is introduced? Do the rotor inlet and exit velocities remain identical? Neither the classical actuator-disk formulation nor the annular actuator-disk extensions of Sharpe and Wood introduce physically distinct inlet and exit velocities at the rotor plane.

Even when swirl is considered, the axial induction factor $a$ is usually taken from the original derivation (GMT) which yielded $a = 1/3$. Alternatively, it can be assumed to have a fixed radial distribution based on experimental data where its value is treated as an arbitrary parameter that is fine-tuned to match the observed performance data or to satisfy empirical relations derived from wind tunnel tests or field data. However, experimental testing requires that the blades be first built, which somewhat defeats the purpose of the analytical prediction when budgets are limited. As noted by Bourhis *et al.* (2023), the choice of $a$ is often arbitrary, unless of course more detailed models such as Blade Element Momentum Theory (BEMT) or CFD are used and $a$ is obtained as part of the numerical solution.

The contention of this paper is that once consideration of swirl is incorporated into the analysis, there is no longer any reason to hold to the value of $a = 1/3$—an artifact of the initial derivation that ignored swirl. But if not $a = 1/3$, some new assumption must be made.

The most optimistic assumption that can be made is that there are no losses through the blade row, and that the magnitude of the velocity vector returns to its free-stream value as it reaches the "actuator disc" exit plane (but with a lower pressure). As shown herein, this assumption results in the value $a = 0.18$ at peak efficiency. Studies measuring induction factors on built wind turbines in wind tunnels report $(0.125 \le a \le 0.42)$ for a tip-speed ratio of 5 (Neff and Meroney 1997), $(0.122 \le a \le 0.165)$ at low tip-speed ratio (Lindenburg 2003), $(0.2 \le a \le 0.34)$ over blade span at tip-speed ratio of 9 (Fritz *et al.* 2024), and $(0.1 < a < 0.45)$ for tip-speed ratios from 1.5 to 7.5 (Larwood 2001).

For discussion of some inefficiencies of wind turbines which are not discussed here, see Georgiou and Theodoropoulos (2011).

## 2. Analysis

Betz's 1928 paper imagined a design wherein drag is minimized to a negligible amount, and the torque of the wind turbine is produced entirely by lift rather than drag. Referring to Figure 1, we see that the power extracted from the wind turbine is

$$\dot{W} = \frac{\dot{m}}{2}({v_\infty}^2 - {v_3}^2) \,, \tag{2}$$

where the mass flow rate $\dot{m}$ is

$$\dot{m} = \rho \, v_N \, A \,, \tag{3}$$

$A$ is any section of the overall wind turbine streamtube, and $v_N$ is the component of air velocity normal to $A$. The velocity just upstream of the wind turbine is $v_1$, and that just downstream of it is $v_2$. Also, $A_1 \approx A_2$ is the circular area swept by the blades.

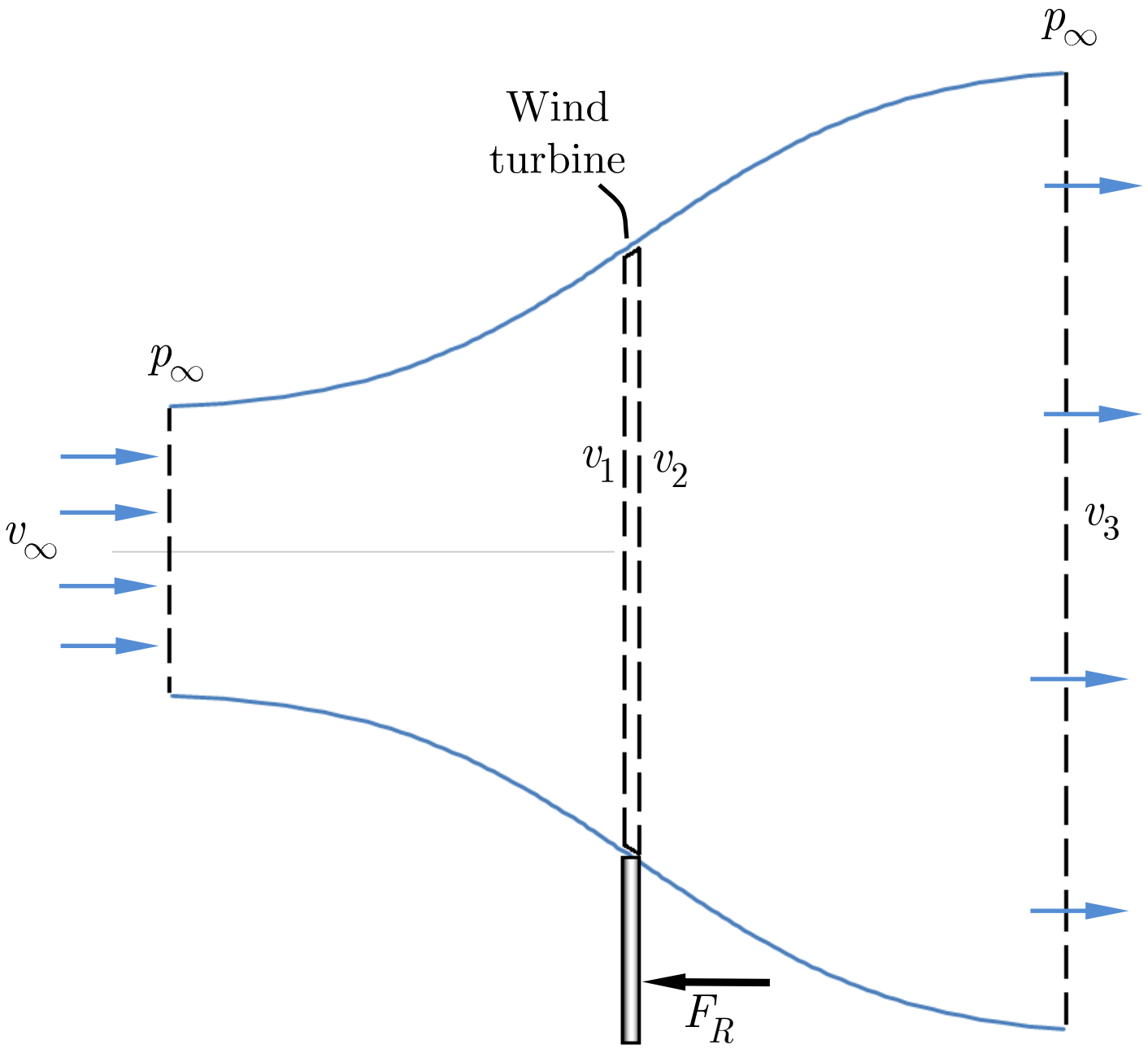

Figure 1. Control volumes used in analysis of wind turbine.

Apply the momentum equation to the overall control volume in the figure, remembering to include the reaction force $F_R$ at the wind turbine support, and obtain:

$$F_R = \dot{m}(v_3 - v_\infty)\,. \tag{4}$$

Note that $p_3 = p_\infty$, so these pressure terms dropped out.

Likewise, apply the momentum equation to the small control volume to obtain

$$F_R \;+\; p_1 A_1 \;-\; p_2 A_2 \;=\; \dot{m}(v_{N2} - v_1)\,.$$

For the axial velocity component $v_{N2}$ of $v_2$, we can say that $v_{N2} = v_1$, so the reaction force in the streamwise direction simplifies to

$$F_R \;=\; (p_2 - p_1)A\,. \tag{5}$$

Now eliminate $F_R$ from (4) and (5), and use

$$\dot{m} = \rho\, v_1\, A \tag{6}$$

to obtain

$$\rho v_1 (v_\infty - v_3) \;=\; (p_1 - p_2)\,. \tag{7}$$

This is a useful result that will be used to replace pressures with velocities.

Now apply Bernoulli's equation between the upstream (freestream) plane and plane 1:

$$p_\infty + \frac{1}{2}\rho {v_\infty}^2 \;=\; p_1 + \frac{1}{2}\rho {v_1}^2\,, \tag{8}$$

and also apply Bernoulli's equation between planes 2 and 3 to obtain:

$$p_2 + \frac{1}{2}\rho {v_2}^2 \;=\; p_3 + \frac{1}{2}\rho {v_3}^2\,. \tag{9}$$

At this point, Betz (and presumably Lanchester and Joukowsky, if not explicitly) made the assumption that

$$v_2 \approx v_1\,. \qquad \text{(Betz)} \qquad (10)$$

This allowed him to use (8) and (9) (eliminating $p_3$ and $p_\infty$) to obtain

$$p_2 - p_1 \;=\; \frac{1}{2}\rho({v_3}^2 - {v_\infty}^2)\,. \qquad \text{(Betz)} \qquad (11)$$

However, due to swirl at plane 2, we could write ${v_2}^2 = {v_{\theta 2}}^2 + {v_{N2}}^2$ in (9), where $v_{\theta 2}$ is the swirl component of velocity at the exit plane. The assumption $v_2 = v_1$ causes a quandary. Because the primary function of the turbine—the generation of torque—requires swirl, admitting swirl requires the velocity decomposition:

$$ {v_2}^2 = {v_{\theta 2}}^2 + {v_{N2}}^2\,. \qquad (12)$$

So if $v_2 = v_1$, then ${v_{N2}}^2 = {v_1}^2 - {v_{\theta 2}}^2$, which means that either

$$v_{N2} < v_1 \qquad \text{or} \qquad v_{\theta 2} = 0\,. \qquad (13)$$

Consequently, the condition $v_2 = v_1$ can only be satisfied by either:
(1) $\left|v_{\theta 2}\right| > 0$, which allows for torque but requires $v_{N2} < v_1$, violating continuity, or
(2) $v_{\theta 2} = 0$, implying no swirl and therefore no torque production.

But torque extraction requires a non-zero swirl velocity component $v_{\theta 2}$ at plane 2, and the condition $v_{N2} = v_1$ must also hold, by continuity. The following discussion explains what reasoning can be used to allow for this.

Regardless of the how the wind turbine is designed, it will present some obstruction to the free flow of air, so the velocity at plane 1 will be lower than $v_\infty$. And there is no way for the turbine to accelerate the air above $v_\infty$, absent some sort of funneling structure. And since the blades are unconfined by any sort of cowl, this is unlikely. As air passes through the wind turbine, the effect of the blades is merely to turn the flow. As the air emerges to plane 2, the upper-bound case is obtained by assuming reversible recovery of the velocity magnitude to its original $v_\infty$. So for an upper limit, the following condition is proposed:

$$v_2 = v_\infty\,. \qquad \text{(better)} \qquad (14)$$

(But the pressure at plane 2 will be lower than at plane 1 due to the extraction of energy by the blades.) This assumption is the equivalent of assuming zero loss through the rotor.

Using this limit condition in (9) allows us to calculate the following pressure difference that is consistent with both continuity and angular momentum, unlike (11):

$$p_1 - p_2 \;=\; \rho {v_\infty}^2 \;-\; \frac{1}{2}\rho {v_3}^2 \;-\; \frac{1}{2}\rho {v_1}^2\,. \qquad \text{(better)} \qquad (15)$$

Now the results will differ from those of the historical limit, but for completeness, the remainder of the derivation leading to the 59% will be recounted briefly.

Betz factored $(v_3 - v_\infty)$ out of the right side of (11) and combined it with (7) to obtain

$$v_1 \;=\; \frac{1}{2}(v_\infty + v_3)\,. \qquad \text{(Betz)} \qquad (16)$$

Betz then substituted this value of $v_1$ into (6), and the result into (2), to give the following relation for power:

$$\dot{W} \;=\; \frac{1}{2}\rho\,\frac{(v_\infty + v_3)}{2}\,A({v_\infty}^2 - {v_3}^2)$$

$$= \frac{1}{4}\rho A {v_\infty}^3 \left(1 + \frac{v_3}{v_\infty}\right)\left[1 - \left(\frac{v_3}{v_\infty}\right)^2\right] \qquad \text{(Betz)} \qquad (17)$$

Taking the derivative of $\dot{W}$ with respect to $v_3/v_\infty$ and setting the result to zero indicates that the power will be greatest when the design is such that

$$\frac{v_3}{v_\infty} = \frac{1}{3}. \qquad \text{(Betz)} \qquad (18)$$

Incidentally, this, together with (16), yields an axial induction factor $a$ of

$$a = \frac{(v_\infty - v_1)}{v_\infty} = \frac{1}{3}.$$

Substituting (18) into (17), we compute that the maximum available power from the wind is

$$\dot{W}_{\max} = \frac{16}{27}\frac{\rho}{2} A {v_\infty}^3. \qquad \text{(Betz)} \qquad (19)$$

Therefore, the so-called *Betz limit* says that, at most,

$$\frac{16}{27} \approx 59.3\% \qquad \text{(Betz)}$$

of the oncoming kinetic energy can be converted to useful energy.

As mentioned earlier, we can improve on this estimate, however. We return now to the more accurate pressure difference in (15) and eliminate the pressure terms there, as before, by combining with (7) to obtain the following alternative to (16):

$$v_1 = v_3 - v_\infty + \sqrt{v_\infty(3v_\infty - 2v_3)}\,. \qquad (20)$$

Incidentally, if the assumption (14) is not made, this relation would be

$$v_1 = v_3 - v_\infty + \sqrt{2(v_\infty - v_3)v_\infty + {v_2}^2}\,. \qquad (21)$$

Substitute (20) into (6), and the result into (2). This gives the following alternative to (18):

$$\dot{W} = \frac{\rho A}{2}[v_3 - v_\infty + \sqrt{v_\infty(3v_\infty - 2v_3)}]({v_\infty}^2 - {v_3}^2)\,.$$

To find a value for $v_3$ that maximizes power, we could differentiate this power with respect to $v_3$, similarly to the strategy of Betz. But the differentiated expression for $\partial \dot{W}/\partial v_3$ becomes too unwieldy. So instead, we can solve for $v_3$ numerically by maximizing power using a spreadsheet solver, and obtain:

$$v_3 = 0.2182 v_\infty\,, \qquad (22)$$

which gives a maximum power of

$$\dot{W}_{\max} = 0.7803 \frac{\rho}{2} A {v_\infty}^3. \qquad \text{(better)} \qquad (23)$$

Therefore, at most 78% (not the 59% predicted by (19)) of the energy in the on-coming air stream can be converted into useful work by the turbine.

Substituting (22) into (20) gives the velocity at the turbine inlet plane $v_1$:

$$v_1 = 0.8193\, v_\infty\,. \qquad (24)$$

This yields an axial induction factor $a$ of

$$a = \frac{(v_\infty - v_1)}{v_\infty} \approx 0.18\,. \tag{25}$$

Recall that the Lanchester-Betz analysis yielded the value $a = 1/3$.

Substituting (22) and (24) into (7) gives the following relation for the pressure drop across the wind turbine:

$$p_1 - p_2 = 0.6405\,\rho\, {v_\infty}^2\,. \tag{26}$$

The thrust at the support structure can be calculated by $F_R = \dot{m}(v_{2N} - v_\infty)$, where $v_{2N} = \sqrt{{v_2}^2 - {v_{2\theta}}^2}$. Thus

$$F_R = \dot{m}\left(\sqrt{{v_2}^2 - {v_{2\theta}}^2} - v_\infty\right),$$

and by (14),

$$F_R = \dot{m} v_\infty \left[\sqrt{1 - \left(\frac{v_{2\theta}}{v_\infty}\right)^2} - 1\right]. \tag{27}$$

## 3. Efficiency Versus Blade Speed

The split between axial and swirl velocity at plane 2 depends on the amount of blade turning. If turning is minimal, then we are not extracting as much energy as we could. If turning is maximal, then there is no axial air velocity component, and hence no through flow. In the latter case, power produced would be zero according to (2). Blade twist determines the swirl component $v_{\theta 2}$ of the velocity at plane 2 which, in turn, depends on radius. So consider all of the above equations to be applied to a stream-annulus rather than the entire flow area. For this scenario, we go back and make the following reconsideration:

$$W \rightarrow \mathrm{d}W\,, \qquad \dot{m} \rightarrow \mathrm{d}\dot{m}\,.$$

Then, the new differential form of (2), is

$$\mathrm{d}\dot{W} = \mathrm{d}\dot{m}\,\frac{({v_\infty}^2 - v_3^2)}{2}\,. \tag{28}$$

Another equation for power, in addition to this relation, can be written by using the product of blade rotation rate $\Omega$ and the torque $\mathrm{d}T$ associated with the annulus of the blade passage. The torque $\mathrm{d}T$ created by the air in this annulus is given by the angular momentum equation

$$\mathrm{d}T = \mathrm{d}\dot{m}\, v_{\theta 2}\, r\,. \tag{29}$$

This affects blade twist because of the presence of the swirl velocity $v_{\theta 2}$, but before considering blade twist, it is best to focus solely on the turning of the flow. Therefore, we take a step back and conceptually unwrap the annulus so that it is just a rectangular opening normal to an airflow, and within which a blade is simply translating, with velocity $v_{\theta 2}$, transversely to, and as a reaction to, the wind. Now the question to examine is: How much should the translating blade turn the flow in order to capture the most wind energy possible?

First, equate the power $\mathrm{d}\dot{W} = \Omega\,\mathrm{d}T$ associated with (29) to that in (28):

$$\mathrm{d}\dot{W} = \mathrm{d}\dot{m}\,\frac{({v_\infty}^2 - v_3^2)}{2} = \Omega\,\mathrm{d}T = \Omega\,\mathrm{d}\dot{m}\, v_{\theta 2}\, r\,. \tag{30}$$

Solving for the velocity $v_{\theta 2}$ gives:

$$v_{\theta 2} = \frac{{v_\infty}^2 - {v_3}^2}{2U}. \tag{31}$$

Substitute this into (12), and solve for $v_{N2}$, which, by continuity, is also $v_1$. This gives

$$v_1 = \sqrt{{v_2}^2 - \left(\frac{{v_\infty}^2 - {v_3}^2}{2U}\right)^2}.$$

Equating this relation to (20) and using (14) gives

$$2U\left[v_3 - v_\infty + \sqrt{v_\infty(3v_\infty - 2v_3)}\right] = \sqrt{(2Uv_\infty)^2 - ({v_\infty}^2 - {v_3}^2)^2} \tag{32}$$

Solving for $U$ gives the local blade speed as a function of velocity $v_3$ far downstream:

$$\frac{U}{v_\infty} = \frac{1-(v_3/v_\infty)^2}{2\sqrt{1-\left[(v_3/v_\infty)-1 + \sqrt{3-2(v_3/v_\infty)}\right]}} \qquad (v_2 = v_\infty) \tag{33}$$

This is plotted in Figure 2 with $U/v_\infty$ on the horizontal axis.

Importantly, (33) has no valid solutions for blade speed outside the range:

$$\left(\frac{1}{2\sqrt{1-(\sqrt{3}-1)^2}} \leq \frac{U}{v_\infty} < 1\right). \tag{34}$$

This can be verified by substituting the values $v_3/v_\infty = 0$ and $v_3/v_\infty = 1$ into (33). (A second solution branch does exist for higher blade-speed tip ratios, but this requires $v_2 < v_\infty$ and will be discussed later.) The reason for the absence of solutions of (33) for lower blade tip ratios lies in (30), which assumes that all available power in (28) is used to produce power that turns the turbine. For extremely low blade speed ratios, this assumption cannot be satisfied because even though a slowly moving blade can turn the flow in the opposite direction more easily than a blade moving rapidly, the problem is that for a wind turbine, turning the air in the required direction is not the only requirement. The blade must also exchange enough angular momentum to extract that energy, and a slowly moving blade cannot do that as well. According to (31), the energy extracted per unit mass is

$$\Delta e = \Omega r\, v_{\theta 2} = U(r)\, v_{\theta 2}.$$

If the blade speed $U$ is small, then to extract a given amount of energy, $v_{\theta 2} = \Delta e/U$ must become large. But $v_{\theta 2}$ is capped by the Bernoulli constant associated with the freestream conditions, even if the static pressure were to drop to zero. Some of these impossible values (corresponding to negative far-field velocity $v_3/v_\infty$) are shown with short dashed curves in Figure 2. In Table 1, only values corresponding to positive $v_3/v_\infty$ are tabulated.

The efficiency, also plotted in Figure 2 and tabulate in Table 1, can be computed by taking the ratio of (28) to the total power in the oncoming wind in an equal area, namely

$$\mathrm{d}\dot{W}_{\max} = (\rho/2){v_\infty}^3\,\mathrm{d}A. \tag{35}$$

For example, the value of $v_3$ given by (22) appears in the column for $v_3/v_\infty$ in the row of maximum efficiency. According to Table 1, the dimensionless blade speed that gives the peak efficiency of 78% is around $U/v_\infty \approx 0.83$, but the entire blade cannot see this efficiency since the blade must rotate as a rigid body, with higher speeds near the tip and lower speeds near the root.

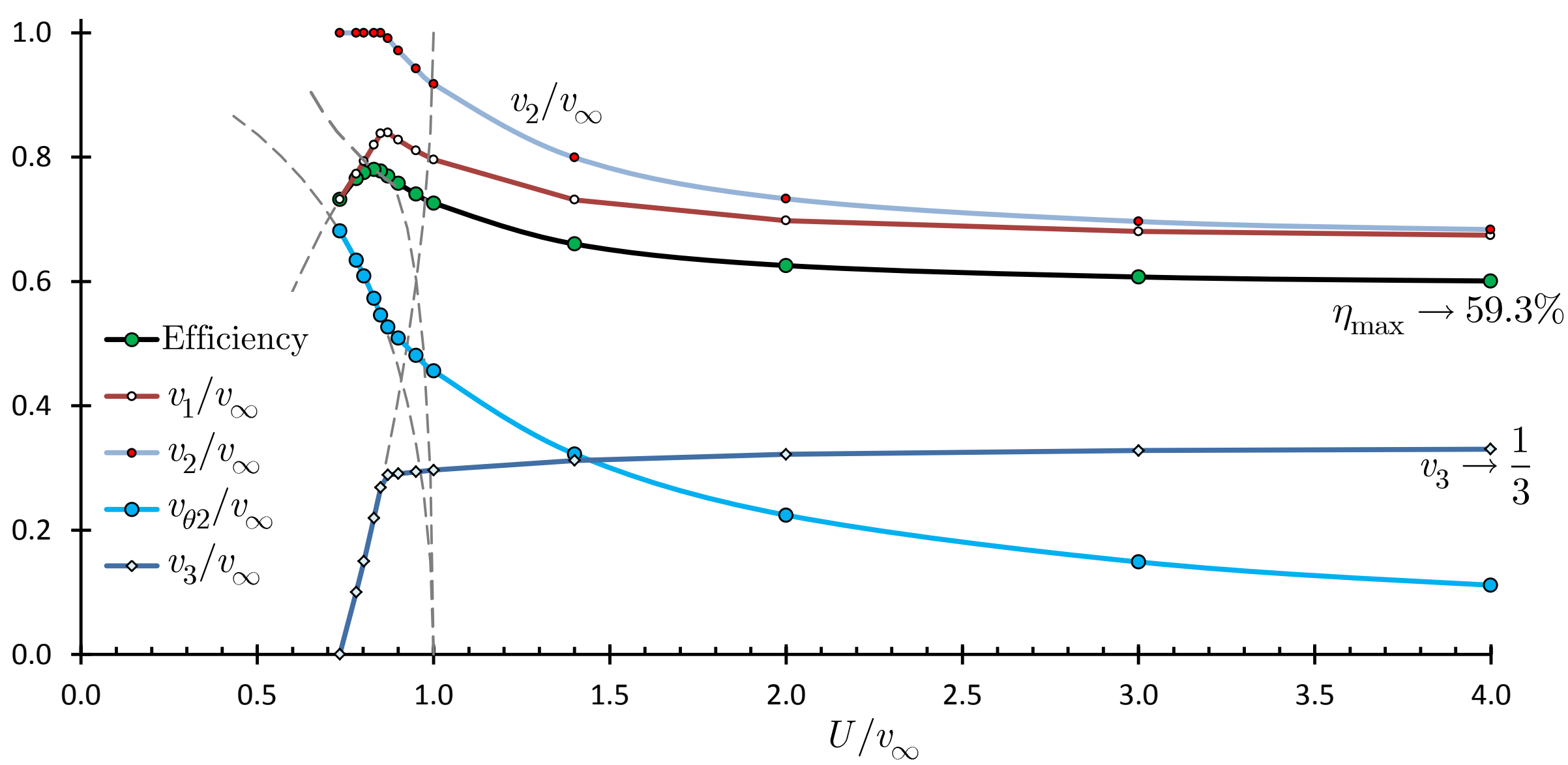

Figure 2. Efficiency and dimensionless velocities versus dimensionless blade speed $U/v_\infty$.
Short-dashed curves: set $v_2 = 1$ (use (20) and (33)).
Long-dashed curves: allow $v_2 < 1$ (use (21) and (37)).

| Efficiency | $U/v_\infty$ | $v_1/v_\infty$ | $v_{\theta 2}/v_\infty$ | $v_2/v_\infty$ | $v_3/v_\infty$ | RPM for $v_\infty = 14$ m/s | | | | | |
|---|---|---|---|---|---|---|---|---|---|---|---|
| | | | | | | 20 | 100 | 200 | 500 | 1000 | 2000 |
| | | | | | | Local Blade Radius [m], $v_\infty(U/v_\infty)/\Omega$ | | | | | |
| 0.594 | 10 | 0.668 | 0.044 | 0.669 | 0.333 | 66.85 | 13.37 | 6.68 | 2.674 | 1.337 | 0.668 |
| 0.594 | 9 | 0.668 | 0.049 | 0.670 | 0.333 | 60.16 | 12.03 | 6.02 | 2.406 | 1.203 | 0.602 |
| 0.595 | 8 | 0.669 | 0.056 | 0.671 | 0.333 | 53.48 | 10.70 | 5.35 | 2.139 | 1.070 | 0.535 |
| 0.595 | 7 | 0.669 | 0.064 | 0.672 | 0.332 | 46.79 | 9.36 | 4.68 | 1.872 | 0.936 | 0.468 |
| 0.596 | 6 | 0.670 | 0.074 | 0.674 | 0.332 | 40.11 | 8.02 | 4.01 | 1.604 | 0.802 | 0.401 |
| 0.598 | 5 | 0.672 | 0.089 | 0.678 | 0.331 | 33.42 | 6.68 | 3.34 | 1.337 | 0.668 | 0.334 |
| 0.601 | 4 | 0.674 | 0.111 | 0.684 | 0.330 | 26.74 | 5.35 | 2.67 | 1.070 | 0.535 | 0.267 |
| 0.607 | 3 | 0.680 | 0.149 | 0.697 | 0.328 | 20.05 | 4.01 | 2.01 | 0.802 | 0.401 | 0.201 |
| 0.626 | 2 | 0.698 | 0.224 | 0.733 | 0.322 | 13.37 | 2.67 | 1.34 | 0.535 | 0.267 | 0.134 |
| 0.660 | 1.4 | 0.731 | 0.322 | 0.799 | 0.312 | 9.36 | 1.87 | 0.94 | 0.374 | 0.187 | 0.094 |
| 0.726 | 1 | 0.796 | 0.456 | 0.917 | 0.297 | 6.68 | 1.34 | 0.67 | 0.267 | 0.134 | 0.067 |
| 0.741 | 0.950 | 0.811 | 0.481 | 0.943 | 0.294 | 6.35 | 1.27 | 0.64 | 0.254 | 0.127 | 0.064 |
| 0.758 | 0.900 | 0.828 | 0.509 | 0.971 | 0.291 | 6.02 | 1.20 | 0.60 | 0.241 | 0.120 | 0.060 |
| 0.769 | 0.870 | 0.839 | 0.527 | 0.991 | 0.289 | 5.82 | 1.16 | 0.58 | 0.233 | 0.116 | 0.058 |
| 0.775 | 0.857 | 0.845 | 0.535 | 1 | 0.288 | 5.73 | 1.15 | 0.57 | 0.229 | 0.115 | 0.057 |
| 0.777 | 0.850 | 0.838 | 0.546 | 1 | 0.269 | 5.68 | 1.14 | 0.57 | 0.227 | 0.114 | 0.057 |
| 0.780 | 0.831 | 0.820 | 0.573 | 1 | 0.219 | 5.55 | 1.11 | 0.56 | 0.222 | 0.111 | 0.056 |
| 0.775 | 0.803 | 0.793 | 0.609 | 1 | 0.150 | 5.36 | 1.07 | 0.54 | 0.215 | 0.107 | 0.054 |
| 0.766 | 0.781 | 0.773 | 0.634 | 1 | 0.100 | 5.22 | 1.04 | 0.52 | 0.209 | 0.104 | 0.052 |
| 0.732 | 0.734 | 0.732 | 0.681 | 1 | 0.000 | 4.91 | 0.98 | 0.49 | 0.196 | 0.098 | 0.049 |

Table 1. Fluid velocities and efficiency as a function of blade speed.

At the upper end of the range in (34), which numerically is

$$\left(0.734 \;\leq\; \frac{U}{v_\infty} \;<\; 1\right), \tag{36}$$

a different assumption fails. For $U/v_\infty > 1$, (33) has no solutions because the assumption in (14) cannot be satisfied (the flow cannot return fully to its free-stream value). So for $U/v_\infty > 1$, the assumption $v_2 = v_\infty$ in (14) must be abandoned. In fact, to find the second branch—the unconstrained optima—we should abandon (14) even slightly before $U/v_\infty$ reaches 1. Instead of (33), the more general relation for dimensionless blade speed becomes:

$$\frac{U}{v_\infty} \;=\; \frac{1-(v_3/v_\infty)^2}{2\sqrt{\left(\dfrac{v_2}{v_\infty}\right)^2 - \left[\dfrac{v_3}{v_\infty} - 1 \;+\; \sqrt{2\left(1-\dfrac{v_3}{v_\infty}\right)+\left(\dfrac{v_2}{v_\infty}\right)^2}\right]^2}}\,, \tag{37}$$

and instead of (20), the more general relation for $v_1$ becomes (21). Values in Table 1 for higher blade speeds will be computed from these more general relations while optimizing for efficiency. This optimization process will reveal a second branch below which the above branch dips. This will be discussed next.

Non-dimensionalizing as follows:

$$x = \frac{v_1}{v_\infty}\,, \qquad y = \frac{v_2}{v_\infty}\,, \qquad z = \frac{v_3}{v_\infty}\,, \qquad s = \frac{v_{\theta 2}}{v_\infty}\,, \qquad u = \frac{U}{v_\infty}$$

allows us write (21) more quickly as

$$x \;=\; z - 1 \;+\; \sqrt{2(1-z)+y^2}$$

Isolating the radical, squaring both sides, and solving for $y^2$ gives

$$y^2 \;=\; x^2 + z^2 - 2xz + 2x - 1 \tag{38}$$

Substitute this into (12), re-written as

$$s^2 = y^2 - x^2\,, \tag{39}$$

to obtain

$$s^2 \;=\; z^2 - 2xz + 2x - 1\,.$$

This factors into

$$s^2 \;=\; (1-z)(2x-1-z)$$

Equate this to the square of (31)

$$s = \frac{1-z^2}{2u} \tag{40}$$

to obtain

$$(1-z)(2x-1-z) \;=\; \frac{(1-z^2)^2}{4u^2}$$

For the non-trivial branch, divide by $(1-z)$, which corresponds to zero energy extraction, and solve for $x$:

$$x \;=\; \frac{1+z}{2} \;+\; \frac{(1-z)(1+z)^2}{8u^2}\,. \tag{41}$$

The advantage of this form over (37) is that $y$ has been eliminated without requiring the reversible condition $y = 1$, and the nested square root has been eliminated. This will be important for differentiating the efficiency to find a maximum.

The efficiency found by taking the ratio of (28) to (35) is:

$$\eta = \frac{\rho v_1({v_\infty}^2 - {v_3}^2)\mathrm{d}A}{2(\rho/2){v_\infty}^3\mathrm{d}A} = \frac{v_1({v_\infty}^2 - {v_3}^2)}{{v_\infty}^3} = x(1-z^2) \tag{42}$$

Substitute (41) into this relation, and obtain

$$\eta = (1-z^2)\left[\frac{1+z}{2} + \frac{(1-z)(1+z)^2}{8u^2}\right]$$

or

$$\eta = (1-z^2)(1+z)\left[\frac{4u^2+(1-z)(1+z)}{8u^2}\right]$$

$$= \frac{(1-z)(1+z)^2(4u^2+1-z^2)}{8u^2}$$

Now locate the extrema:

$$\frac{\partial\eta}{\partial z} = \frac{2(1-z)(1+z)(4u^2+1-z^2)}{8u^2} - \frac{(1+z)^2(4u^2+1-z^2)}{8u^2} - \frac{2z(1+z)^2(1-z)}{8u^2} = 0$$

or

$$(1+z)\left[2(1-z)(4u^2+1-z^2)-(1+z)(4u^2+1-z^2)-2z(1+z)(1-z)\right] = 0$$

Of the two factors that make this zero, the factor $(1+z)$ corresponds to an unrealistic negative velocity $v_3$, so it is discarded. Simplifying the terms inside the brackets gives

$$\left[5z^3 - z^2 - 5z + 1 + (4-12z)u^2\right] = 0\,.$$

This cubic defines the unconstrained optimum ($y$ need not equal 1) at higher blade speeds than the earlier inboard optimum constrained by $y = 1$ $(v_2 = v_\infty)$. Solving for $u^2$ gives a relation for the dimensionless blade speed for this interior branch of maximum efficiency at the higher dimensionless blade speeds:

$$u^2 = \frac{5z^3 - z^2 - 5z + 1}{4(3z-1)}\,.$$

The numerator can be factored into

$$u^2 = \frac{(5z-1)(z-1)(z+1)}{4(3z-1)}\,. \tag{43}$$

The behavior of the numerator is not as complex as it appears at first glance because within the valid range $(0 < z < 1)$, two of the factors in the numerator satisfy

$$(z-1) < 0 \qquad \text{and} \qquad (z+1) > 1\,.$$

So the sign of the numerator is determined entirely from $5z-1$. For $(0 < z < 1/5)$, we have

$$(z-1) < 0\,, \qquad (z+1) > 1\,, \qquad (5z-1) < 0\,,$$

so the numerator is always positive. There is a sign change in the denominator at $z = 1/3$, but below that, $(3z-1) < 0$, so the denominator is always negative. Therefore, $u^2 < 0$, and we discard this interval.

Checking in the range $(1/5 < z < 1/3)$, we see that the numerator is negative, as is the denominator. Therefore, $u^2 > 0$.

For the range $(1/3 < z < 1)$, the numerator remains negative, but the denominator becomes positive, so $u^2 < 0$.

So in the physically possible range $(0 < z < 1)$, the only interval giving a real, nonzero $u$ is:

$$\frac{1}{5} < z < \frac{1}{3}. \tag{44}$$

In summary, (41) contains both branches along which efficiency optima exist. The inboard branch is found by imposing the constraint $y = 1$, and the outboard branch is found imposing the condition by $\partial\eta/\partial z = 0$.

**Transition Point**

The next objective is to locate the point at which the max-efficiency design switches from the inboard curve to the out-board curve. Ironically, the outboard max-efficiency curve is, mathematically, an interior optimum branch because it is an unconstrained optimum ($y$ need not equal 1) found by the condition

$$\frac{\partial\eta}{\partial z} = 0, \tag{45}$$

whereas the inboard max efficiency curved is a boundary optimum, or constrained optimum, because it is constrained by the reversible recovery condition

$$y = 1. \tag{46}$$

The transition from one branch to the other occurs where the unconstrained optimum on the interior branch first reaches the boundary; therefore the two conditions (45) and (46) hold simultaneously. The point at which the two maximum efficiency curves intersect occurs where the interior branch satisfies $y = 1$. Applying the condition in (46) changes (39) to

$$x^2 + s^2 = 1.$$

Substituting (41) and the square of (40) into the above equation gives

$$\left[\frac{1+z}{2} + \frac{(1-z)(1+z)^2}{8u^2}\right]^2 + \frac{(1-z^2)^2}{4u^2} = 1.$$

To apply the condition in (45), we substitute (43) for $u^2$ into the above relation and obtain

$$\left[\frac{1+z}{2} - \frac{(1+z)(3z-1)}{2(5z-1)}\right]^2 + \frac{(1+z)(z-1)(3z-1)}{5z-1} = 1$$

or

$$\left[\frac{z(z+1)}{5z-1}\right]^2 + \frac{(5z-1)(3z^3-z^2-3z+1)}{(5z-1)^2} - 1 = 0.$$

Simplifying gives

$$8z^4 - 3z^3 - 19z^2 + 9z - 1 \;=\; 0\,. \tag{47}$$

The roots of this polynomial can be found numerically to be

$$z \;=\; -1.5919,\; 0.18233,\; 0.28770,\; \text{and}\; 1.49684$$

But only $z = 0.28770$ falls within the allowable range found in (44), so the transition from one optimum branch to the other occurs where

$$z \;=\; \frac{v_3}{v_\infty} \;=\; 0.2877\,. \qquad \text{(at transition)} \tag{48}$$

Substituting this value into (43) gives the dimensionless blade speed at which the transition between the two branches occurs:

$$u \;=\; \frac{U}{v_\infty} \;=\; 0.85713\,. \qquad \text{(at transition)}$$

The other values at the transition point between the two optimum branches are:

$$s \;=\; \frac{v_{\theta 2}}{v_\infty} \;=\; 0.535058\,, \quad y \;=\; \frac{v_2}{v_\infty} \;=\; 1\,, \quad \text{and} \quad x \;=\; \frac{v_1}{v_\infty} \;=\; 0.844815\,.$$

This transition point is indicated in Figure 3, which shows the two max-efficiency branches in bold. The figure shows that the maximum possible efficiency is uniformly higher than the historical 59% limit except at very low blade speeds near the hub. The efficiency at the transition point is, according (42),

$$\begin{aligned} \eta \;&=\; x(1-z^2) \;=\; 0.844815(1-0.2872^2) \\ &=\; 0.7749 \end{aligned}$$

which is close to the global maximum seen in Table 1.

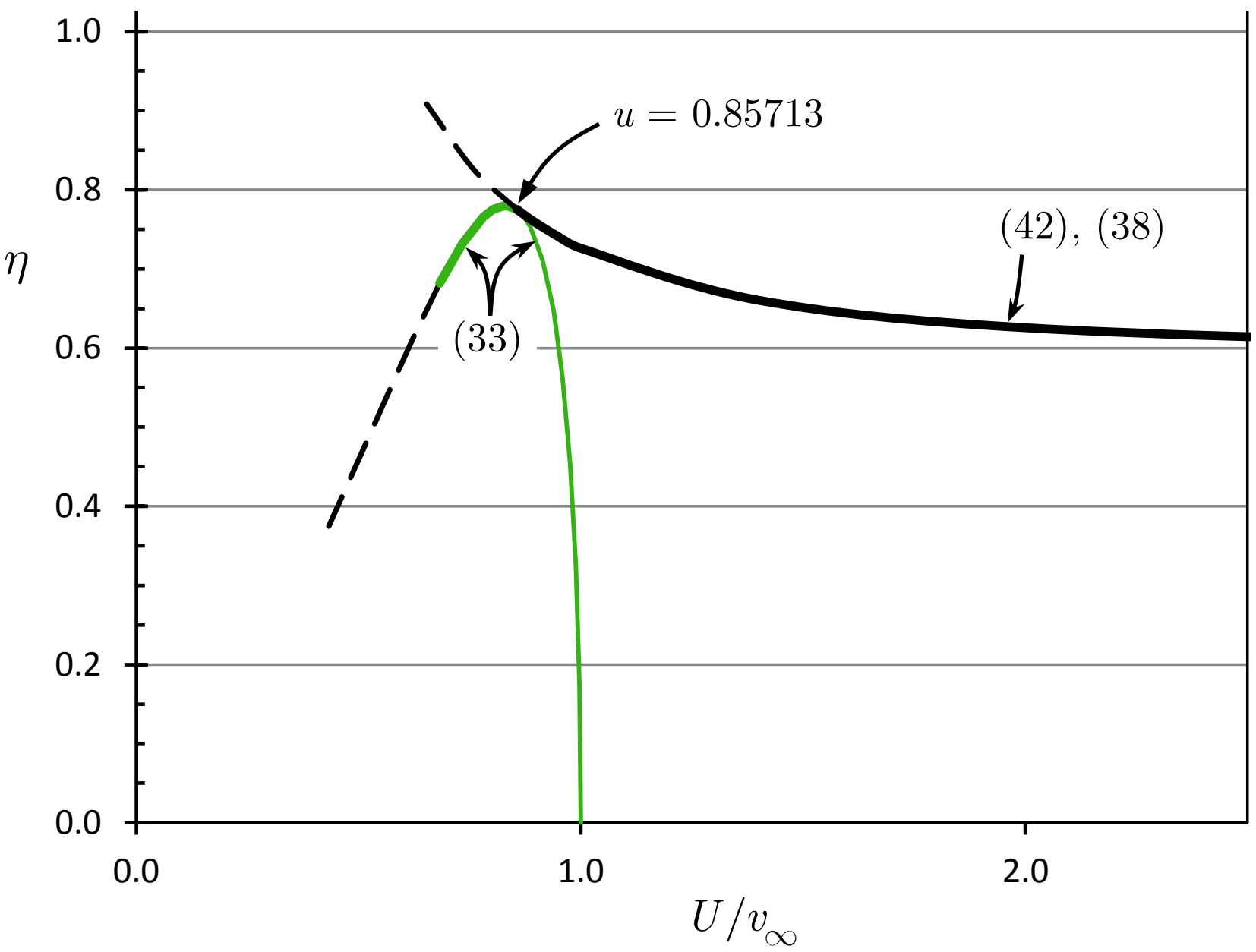

Figure 3. Two branches of maximum efficiency. (Dashed lines: infeasible)

Knowing the efficiency associated with each dimensionless blade speed, we can now translate these blade speeds into radii for a typical wind speed (14 m/s in this case). This can help size the blades. The six columns on the right side of Table 1 show the blade radii implied by different rotation speeds. (The values boxed in the 1000 rpm column will be used in the example below.) Depending upon the tip-to-root ratio of the blade, some inboard area of the annulus may include a region excluded by the lower limit of (36) where no solution exists. In such cases, the wind speed will be too high for this inboard region. This lends credibility to some of the older empirically-designed windmills, such as that shown in Figure 4, that appear to forfeit a substantial inboard portion of the blade span. Also, as $r/r_T$ decreases, the required flow turning grows rapidly. This larger turning requires a larger chord, but such large-chord sections can have poor off-design behavior because they are more sensitive to variations in inflow angle and speed incompatibilities and are more prone to flow separation. A larger $r_R/r_T$ value of wind turbines like that shown in Figure 4 also allows for greater solidity.

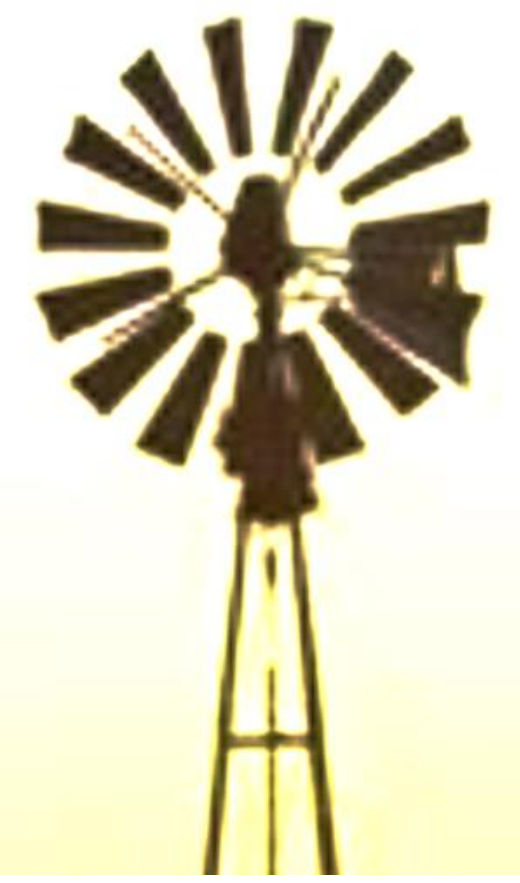

Figure 4. Typical classical windmill design used on farms.

The induction factor $a$ given by (1) is not prescribed as in conventional BEM but emerges from the solution. Figure 5 shows a plot of induction factor based on the values in Table 1. The outboard values shown in the figure are comparable to those reported by Larwood (2001). However, the wind turbine tested by Larwood would not have been designed specifically to the parameters of the present work.

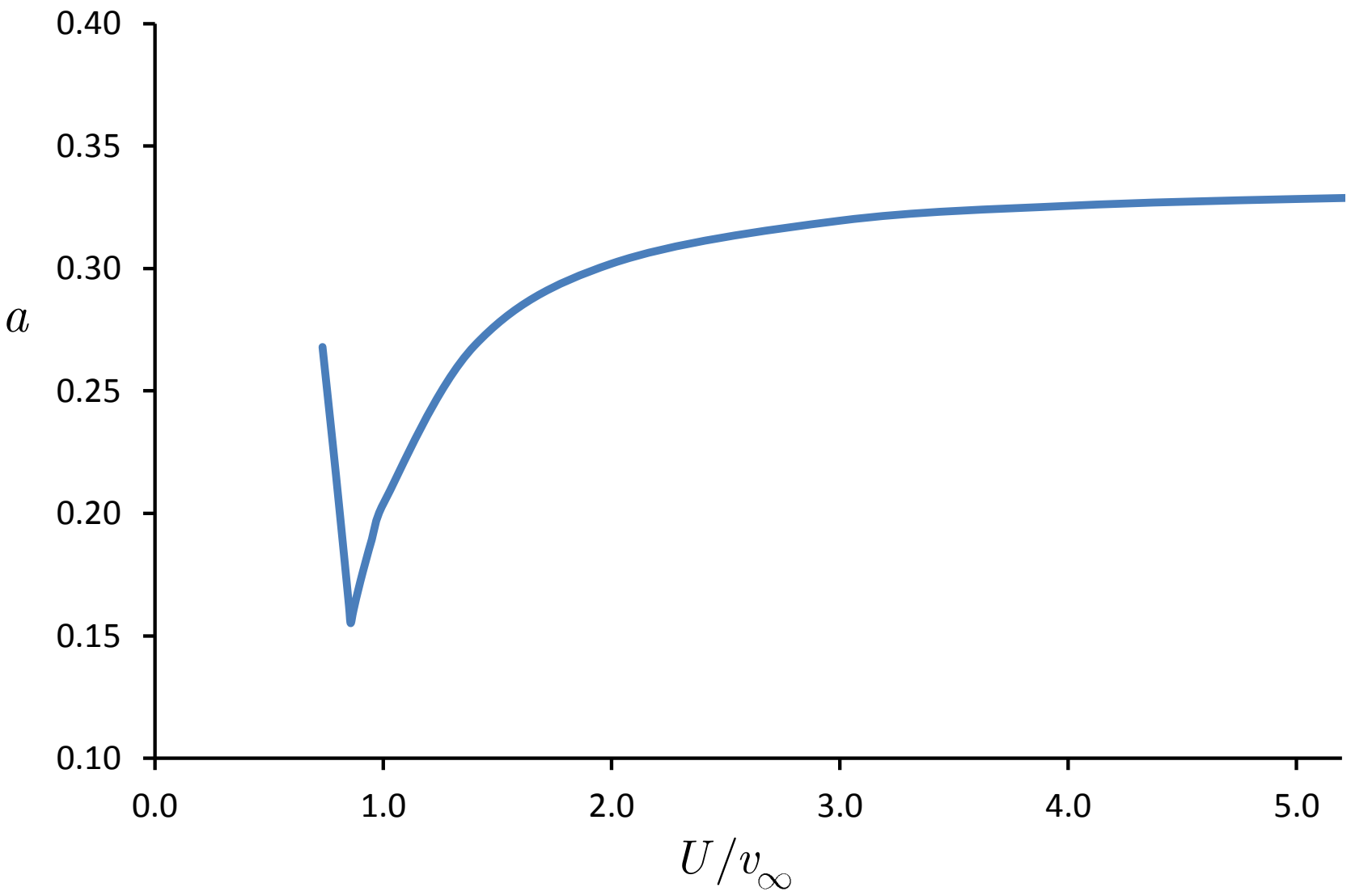

Figure 5. Induction factor $a$ defined by (1) for the example above.

## 4. Pressure at the Wind Turbine Exit Plane

As air passes through the wind turbine, the primary effect of the blades is to turn the flow rather than to slow it down. The swirling flow at the exit plane (plane 2) is much closer to an ideal free vortex than a forced vortex. Wood (2007) observed that the logarithmic singularity due to the $1/r$ dependence is the fundamental problem in considering the effects of swirl on turbine performance and that it arises from ignoring the structure of the hub vortex. But if we restrict our attention to the annulus, we can use the swirl component in Table 1 in the steady momentum equation, written below using Lagrange's decomposition of the acceleration term for convenience:

$$\underbrace{\boldsymbol{\omega}\times\mathbf{v}}_{0\text{ (irrot.)}} + \frac{1}{2}\nabla(\mathbf{v}\cdot\mathbf{v}) = -\frac{\nabla p}{\rho} + \underbrace{\nu\left[\nabla^2\mathbf{v} + \frac{1}{3}\nabla(\nabla\cdot\mathbf{v})\right]}_{0\text{ (inviscid assumption)}},$$

which simplifies to

$$\nabla p = -\frac{1}{2}\rho\nabla({v_\theta}^2).$$

Then

$$\frac{\partial p}{\partial r} = -\frac{1}{2}\rho\frac{\partial}{\partial r}({v_\theta}^2)$$

$$\int_p^{p_T}\mathrm{d}p = -\frac{1}{2}\rho\int_r^{r_T}\mathrm{d}({v_\theta}^2) = -\frac{1}{2}\rho\,{v_\theta}^2\Big|_r^{r_T}$$

Here, the subscript "$T$" refers to the tip of the blade.

$$p - p_R = -\frac{1}{2}\rho\,{v_\infty}^2\left[\left(\frac{v_\theta(r_T)}{v_\infty}\right)^2 - \left(\frac{v_\theta(r)}{v_\infty}\right)^2\right]$$

Figure 6 shows that, by substituting values from the boxed region in Table 1 with $\Omega = 1000\,\text{rpm}$, $r_T = 0.267$ m, and $r_R = 0.098$ m, the pressure drops at the exit plane due to blade twist, and mostly at the root, in qualitative agreement with velocity measurements made by Neff and Meroney (1997).

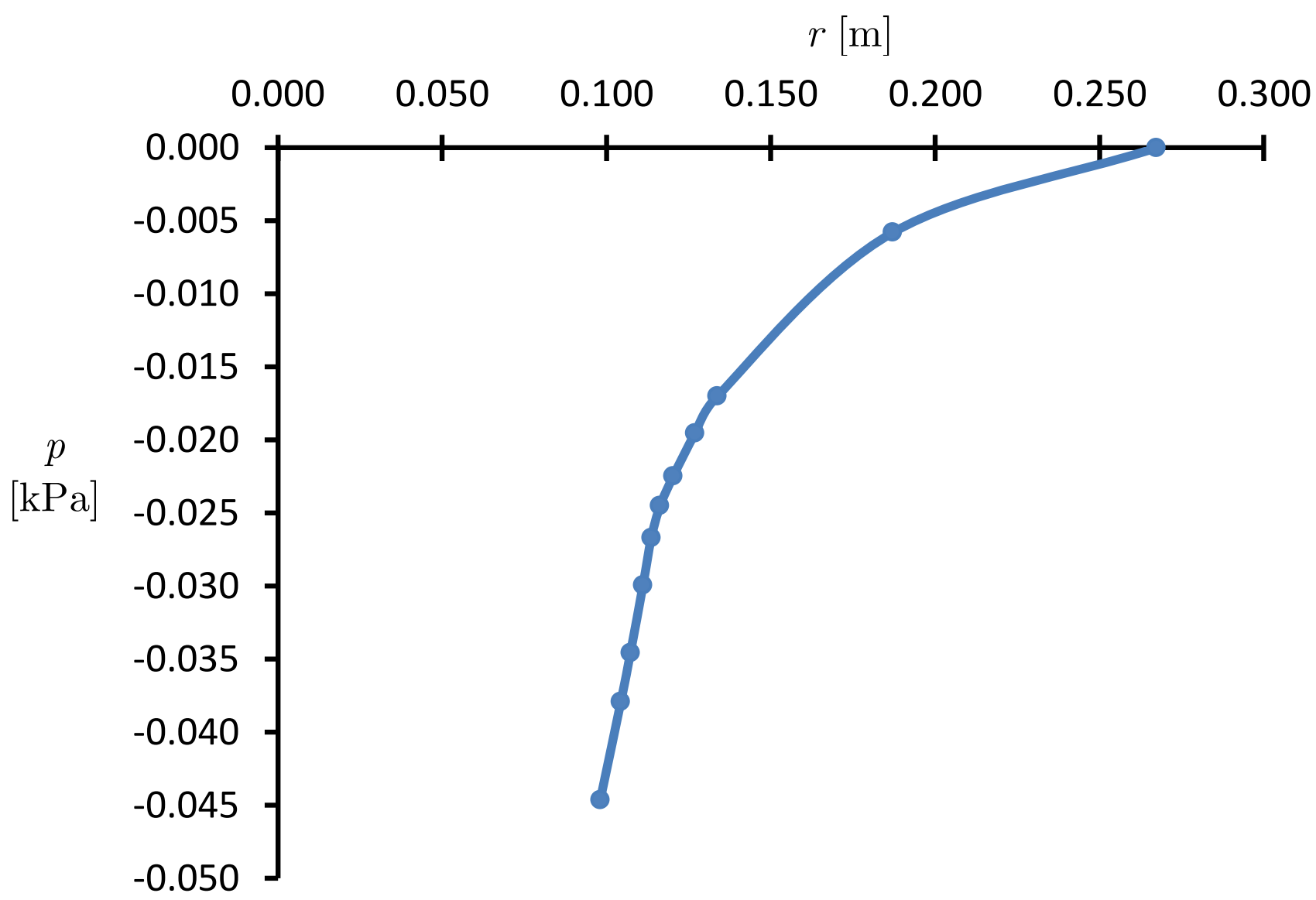

Figure 6. Pressure difference at exit plane relative to free-stream pressure for $v_\infty = 14$ m/s, as a function of dimensionless radius.

## 5. Conclusion

It is remarkable that all three men—Lanchester, Betz, and Joukowsky—while working independently, arrived at the same limit of 59% by making the same assumption $v_1 = v_2$. Although the angular momentum equation (29), which is called Euler's turbine equation, was not needed to derive an upper limit, it must at least be kept in mind in order to avoid the quandary in (13). At the time of Lanchester's work, Euler's turbine equation had already been published in French and German (Bistafa 2021), but it seems to have rarely appeared in English until after 1915 when Lanchester published his limit. This might explain its absence in Lanchester's work. Although Rankine *et al.* (1906, pp. 193-194) and Stodola (1905, p. 28, translation from German) used the concept of Euler's turbine equation and swirl, it was brief and inconspicuous—not highlighted with Euler's name. Other turbine design books written in English before, and just after, Lanchester's 1915 paper (Blaine 1897; Jude 1906; Garnett 1908; Kennedy 1910; Robinson 1910; Holzwarth 1912; and Church 1928) uniformly omitted any concept of Euler's turbine equation, whether named or unnamed. By 1930, Euler's turbine equation began appearing in the English literature (see Glauert 1926, p. 210; Paul 1930; and Kearton 1951, pp. 511, 521. It was not in Kearton's 1922 edition, however.) But it was apparently not associated with the name of Euler until Wislicenus (1947, p. 119-121).

The classical actuator-disk derivation obtains its simplicity by restricting the velocity field to one dimensional and by a closure condition (10). While this assumption leads to a useful upper bound, it does not arise from conservation of angular momentum, as noted in (13), nor from the physical requirements of torque production. It assumes that only pressure will change across the "actuator disc." Even after swirl was incorporated (Glauert 1926, p. 210; 1935, p. 191; Goldstein 1929), the assumption in (10) persisted.

The present work investigates a different idealization in which the rotor exit condition is constrained by angular momentum and reversible recovery. This generates two optimum branches with values higher than the Betz limit. The classical Betz result is recovered in the high blade-speed limit, where the swirl contribution becomes negligible and the assumptions of the one-dimensional formulation become asymptotically valid.

## References

Bergey, K. H., "The Lanchester-Betz limit," *Journal of Energy* **3** (1979) pp. 382-384.

Betz, A., "Das Maximum der theoretisch möglichen Ausnützung des Windes durch Windmotoren," *Zeitschrift fur das Gesamte Turbinenwesen* **26** (1920) pp. 307-309; translated to English in "The Maximum of the Theoretically Possible Exploitation of Wind by Means of a Wind Motor," *Wind Engineering* **37** (2013) pp. 441-446.

Betz, A., "Die Windmühlen im Lichte neuerer Forschung," *Naturwissenschaften* **46** (1927) pp. 905-914; translated in "Windmills in the light of modern research," *NACA-TM-474* (1928).

Bistafa, S. R., "Euler's pioneering general theory of rotating hydraulic machines," *Physics of Fluids* **33** (2021) pp. 101801-1-9.

Blaine, R. G., *Hydraulic Machinery with an Introduction to Hydraulics*, Spon, London (1897).

Bourhis, M., Pereira, M., & Ravelet, F., "Experimental investigation of the effect of blade solidity on micro-scale and low tip-speed ratio wind turbines," *Experimental Thermal and Fluid Science* **140** (2023) pp. 110745-1-14.

Church, E. F., *Steam Turbines*, McGraw-Hill, New York (1928).

Fritz, E., Boorsma, K., & Ferreira, C., "Experimental analysis of a horizontal-axis wind turbine with swept blades using PIV data," *Wind Energy Science* **9** (2024) pp. 1617-1629.

Garnett, W. H. S., *Turbines*, 2nd ed., Bell & Sons, London (1908).

Georgiou, D. P. & Theodoropoulos, N. G., "A momentum explanation for the unsatisfactory Betz model prediction in highly loaded wind turbines," *Wind Energy* **14** (2011) pp. 653-660.

Glauert, H., *The elements of aerofoil and airscrew theory*, Cambridge University Press, Cambridge (1926).

Glauert, H., "Airplane Propellers," in *Aerodynamic Theory: A General Review of Progress*, Vol. 4, ed. by W. F. Durand, Springer, Berlin (1935) pp. 169-360.

Goldstein, S., "On the vortex theory of screw propellers," *Proceedings of the Royal Society of London. Series A* **123** (1929) pp. 440-465.

Greet, R. J., "Maximum windmill efficiency," *Journal of Applied Physics* **51** (1980) pp. 4680-4681.

Holzwarth, *The Gas Turbine*, Griffin & Co., London (1912).

Hütter, U., "Optimum Wind-Energy Conversion Systems," *Annual Review of Fluid Mechanics* **9** (1977) pp. 399-419.

Inglis, D. R., "A windmill's theoretical maximum extraction of power from the wind," *American Journal of Physics* **47** (1979) pp. 416-420.

Joukowsky, N. E., "Vetryak tipa NEJ" ("Windmill of the NEJ Type"), *Transactions of the Central Institute for Aero-Hydrodynamics of Moscow* (1920); also published in *Joukowsky NE. Collected Papers* Vol 6, The Joukowsky Institute for AeroHydrodynamics, Moscow (1937) pp. 405–409 (in Russian).

Jude, A., *The Theory of the Steam Turbine*, Griffin & Co., London (1906).

Kearton, W. J., *Steam Turbine Theory and Practice*, 6th ed. (1951) Pitman & Sons, London.

Kennedy, R., *Steam Turbines: Their Design and Construction*, Whitaker & Co., London (1910).

Lanchester, F. W., "A contribution to the theory of propulsion and the screw propeller," *Transactions of the Institution of Naval Architects* **57** (1915) pp. 98-116.

Larrabee, E. E., “Design of propellers for motorsoarers,” *Science and Technology of Low Speed and Motorless Flight*, NASA Conference Publication 2085, Part I, (1979) pp. 285-303.

Larwood, S., “Wind Turbine Wake Measurements in the Operating Region of a Tail Vane,” presented at the *39th AIAA Aerospace Sciences Meeting, Reno, Nevada, January 8–11* (2001) 10 pp.

Lindenburg, C., “Investigation into Rotor Blade Aerodynamics: Analysis of the stationary measurements on the UAE phase-VI rotor in the NASA-Ames wind tunnel,” NASA Ames wind tunnel campaigns (1999–2001), (2003) 114 pp.

Neff, D. E. & Meroney, R. N., “Mean wind and turbulence characteristics due to induction effects near wind turbine rotors,” *Journal of Wind Engineering and Industrial Aerodynamics* **69-71** (1997) pp. 413-422.

Okulov, V. L. & van Kuik, G. A. M., “The Betz-Joukowsky limit: on the contribution to rotor aerodynamics by the British, German and Russian scientific schools,” *Wind Energy* **15** (2012) pp. 335-344.

Paul, C. S. T., “Propeller type water-turbines,” *Proceedings of the Institution of Mechanical Engineers* **119** (1930) pp. 1409-1421.

Rankine, W. J. M., Millar, W. J., & Donkin, B., *A manual of the steam engine and other prime movers*, 16th ed., Griffin & Co., London (1906).

Rauh, A. & Seelert, W., “The Betz Optimum Efficiency for Windmills,” *Applied Energy* **17** (1984) pp. 15-23.

Robinson, H., *Hydraulic Power and Hydraulic Machinery*, Griffin & Co., London (1910).

Sharpe, D. J., “A General Momentum Theory Applied to an Energy-extracting Actuator Disc,” *Wind Energy* **7** (2004) pp. 177-188.

Stodola, A., *Steam Turbines with an Appendix on Gas Turbines and the Future of Heat Engines*, Van Nostrand, New York (1905).

van Kuik, G. A. M., “The Lanchester-Betz-Joukowsky Limit,” *Wind Energy* **10** (2007) pp. 289-291.

van Kuik, G. A. M., “On the velocity at wind turbine and propeller actuator discs,” *Wind Energy Science* **5** (2020) pp. 855-865.

Wislicenus, G. F., *Fluid Mechanics of Turbomachinery*, McGraw-Hill, New York (1947).

Wood, D. H., “Including Swirl in the Actuator Disk Analysis of Wind Turbines,” *Wind Engineering* **31** (2007) pp. 317–323.